\documentclass[twocolumn,aps,prb,superscriptaddress]{revtex4-2}

\usepackage{graphicx}
\usepackage{amssymb}

\begin{document}

\preprint{preprint}

\title{Magnetic phase diagram of EuPdSn$_2$}


\author{J.G. Sereni}
\affiliation{Low Temperature Division, CAB-CNEA, CONICET, IB-UNCuyo, 8400 Bariloche, Argentina}

\author{I. Čurlik}
\affiliation{Faculty of Sciences, University of Prešov, 17. novembra 1, SK - 080 78 Prešov, Slovakia}

\author{A. Martinelli}
\affiliation{SPIN-CNR, Corso F.M. Perrone 24, 16152 Genova, Italy}

\author{M. Giovannini}
\affiliation{Department of Chemistry, University of Genova, Via Dodecaneso 31, Genova, Italy}

\email[]{jsereni@yahoo.com}

\date{\today}

\begin{abstract}

A magnetic phase diagram for EuPdSn$_2$ is constructed based on magnetization, DC-susceptibility and specific heat as a function of applied field, and neutron diffraction results at zero field. The contribution of the ferromagnetic FM and antiferromagnetic AFM components to the measured magnetic susceptibility can be separated and weighted as a function of the applied field $B$ by taking the temperature dependence of the neutron counts from the FM phase at $B=0$ as a reference. In the magnetically ordered phase, the temperature dependence of the specific heat does not follow the mean-field prediction for the Eu$^{2+}$ $J=7/2$ quantum number rather that for $J=5/2$. This deviation and the consequent discrepancy for the entropy is discussed in the context of a non-Zeeman distribution of the eight fold ground state. 
Comparing the respective variations of magnetization $M(T,B )$ and entropy $S(T,B)$, the 
thermodynamic relationship $\partial M/\partial T = \partial S/ \partial B = 0$ is observed for the ranges: $0<T<10$\,K and $0<B<0.4$\,T.

\end{abstract}

\keywords{Eu copounds, magnetism, phse diagram}

\maketitle


\section{Introduction}

Eu$^{2+}$ ions exhibit a unique property among the rare earth series that allow them to play a 
role in two completely different scenarios because they can mimic the magnetism of $J=7/2$ Gd$^{3+}$ atoms and chemically substitute the divalent alkaline earths such as Sr and Ca. 
This occurs because the localization of a band electron into the Eu$^{3+}$ electronic configuration: [Xe][6s$^2$5d$^1$4f$^6$] leads to the same occupation of the $4f$ orbital as that 
of Gd$^{3+}$:  [Xe][(6s5d)$^2$4f$^7$] for Eu$^{2+}$. As expected, the atomic volume  of Eu$^{2+}$ is considerably increased (from about 24.4\AA$^3$ to 33.4\AA$^3$) since there is an  
addditional electron without electronic charge compensation by another proton.

\begin{figure}
\begin{center}
\includegraphics[width=20pc]{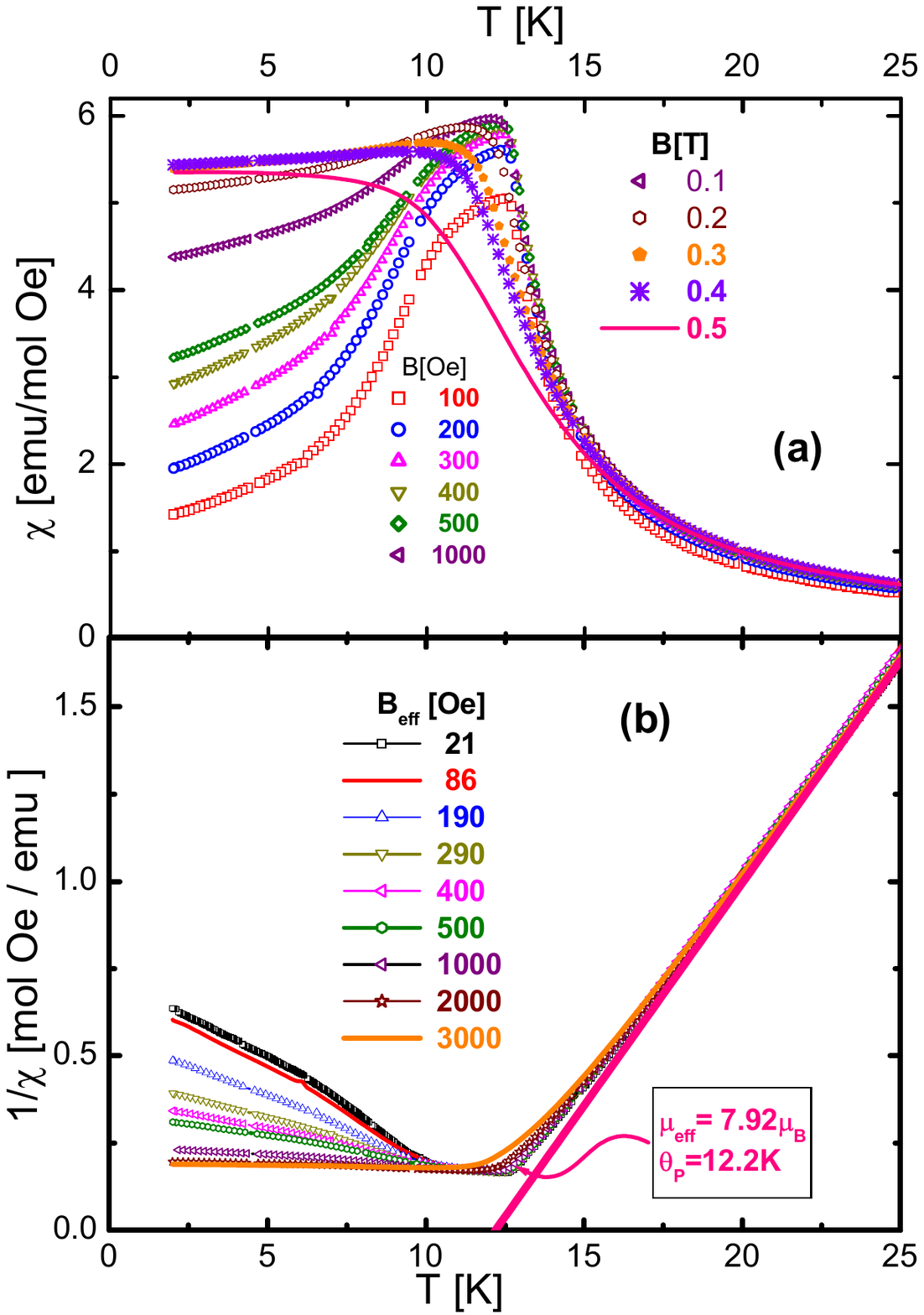}
\caption{(Color online) a) Magnetic susceptibility in the field range: $0.01\leq B \leq 0.5$\,T, after \cite{1-1-2}. For practical reasons, the [Oe] units are used in the lowest range of field ($B\leq 
0.1$\,T). b) Inverse susceptibility according to the effective field $B_{eff}$ (see text), including a measurement at $B_{eff}= 21$\,Oe. Notice the deviation from the pure C-W low approaching 
$T_{ord} $ for $B>2000$\,Oe. \label{F1}}
\end{center}
\end{figure}

There are numerous examples of Eu$^{2+}$ behaving magnetically like Gd$^{3+}$, see for example \cite{EuPtSi3,Seiro,1-1-2,2-2-1} and references therein. Regarding the substitution of 
Eu$^{2+}$ in 2+ 
alkaline earth (e.g. Sr or Ca) sites, the number of publications has rapidly increased because of its  recent applications in optical properties, e.g. to shift activated emissions towards blue spectrum 
\cite{LEDs,CaS:Eu}. 

On the contrary, the application of the magnetic properties of Eu$^{2+}$ 
in biological systems \cite{MCD,MRI} is scarce. However, the fact that magnetic circular dichroism can  be used in these systems according the magnetic nature of Eu$^{2+}$ provides an 
alternative 
microscopic technique that has not yet been fully exploited. Considering this potential application, it is advisable to use the study of magnetic properties of Eu$^{2+}$ compounds as possible  
reference for future investigations of more complex and even biological systems. 

The present work is devoted to the analysis of the evolution of the magnetic properties of EuPdSn$_2$ under magnetic field $B$ with the aim of constructing a magnetic phase diagram. It 
is based on the field-dependent magnetic and thermal measurements \cite{1-1-2}, that takes into account the magnetic phase separation in the ground state of EuPdSn$_2$ revealed by neutron 
diffraction 
\cite{Martinelli} measurements at zero field.

\section{Magnetic properties}

\subsection{Magnetic Susceptility}

The main features reported for this compound at zero field are: an antiferromagnetic (AFM) like cusp at $T_{ord}$=12.5\,K in contrast to the ferromagnetic (FM) molecular field 
suggested by the positive paramagnetic temperature $\theta_P = 11.5$\,K \cite{1-1-2}.   
Coincidentally, this maximum is not a canonical AFM-cusp because it is followed at lower temperatures by a pronounced hump, whose intensity increases with field as seen in Fig.~\ref{F1}
a. As a result the $\chi (T \to 0)$ tail increases significantly with $B$, becoming dominant around $B=0.1$\,T and showing saturation slightly above $B ~ 0.4$T.

It is well known that in low-field studies, the effective applied field $B_{eff}$ can be affected by any residual magnetic field in the magnet when it is also used to produce intense field. This  
variation must be taken into account in low field measurements where the difference between nominal and effective field may become relevant. In fact, in  Fig.~\ref{F1}a it is possible to 
appreciate a deviation of the measured $\chi(T)_{meas}$ at nominal $B=100$\,Oe from other curves at higher field. 
To analyze whether this deviation is intrinsic or due to the mentioned experimental feature, we have tuned the effective $B_{eff}$ values, and merged all $\chi(T,B)$ results into a single curve as required for the pure paramagnetic regime. A further measurement, performed 
at nominal field $B=40$\,Oe, was also included in Fig.~\ref{F1}b as the inverse 
susceptibility  $\chi(T,B)^{-1}$ where the paramagnetic Curie-Weiss CW law is represented by a straight line whose slope corresponds to the inverse of the effective magnetic moment $
\mu_{eff}$. 

\begin{figure}
\begin{center}
\includegraphics[width=20pc]{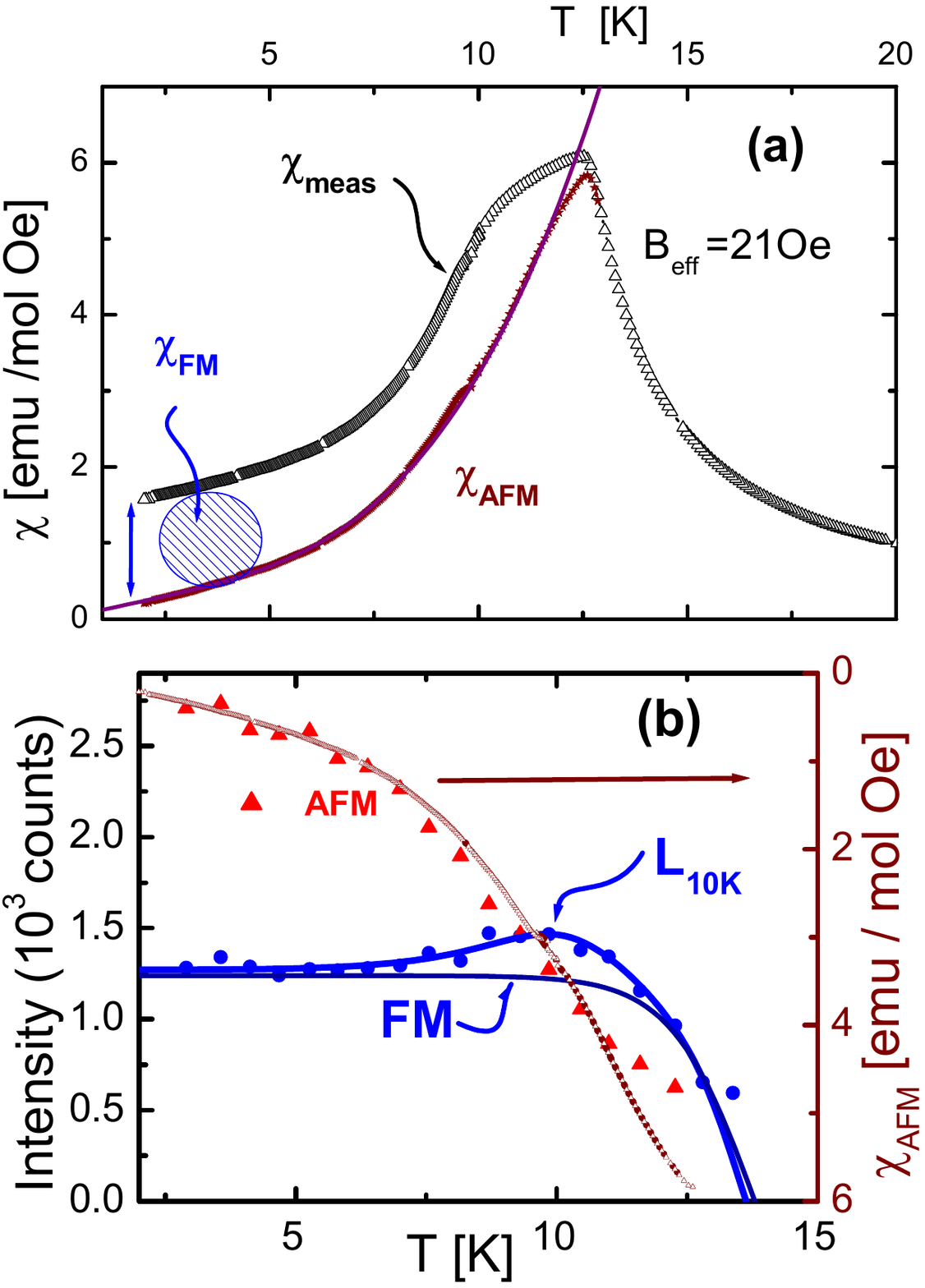}
\caption{(Color online) a) Detachment of FM and AFM contributions from measured $\chi(T)_{meas}$. The blue shadowed circle indicates the $\chi_{FM}$ contribution based on FM neutron 
counts. Continuous brown curve: fit of $\chi_{AFM}(T)$, see the text. b) Temperature dependence of magnetic Bragg peak intensities: FM - blue circles and AFM - red triangles, after 
\cite{Martinelli}. 
Continuous-blue curve: $\chi_{FM} = FM(T)+ L_{10\,K}(T)$, brown-curve comparison between AFM neutron counts (left axis) and $\chi_{AFM}
(T)$ (right axis) after subtracting the FM contribution, notice the reversed values of $\chi_{AFM}(T)$ to allow the comparison.  \label{F2}}
\end{center}
\end{figure}

As it can be seen in the field values depicted in Fig.~\ref{F1}a and b, the difference between $B$ and $B_{eff}$becomes significant at low intensities. This procedure allows us to obtain 
more precise values for the characteristic parameters such as the Curie 
constant: $C_C =7.9$\,emu\,K/mol, which corresponds to $\mu_{eff} =7.92\mu_B$, and  $\theta_P= 12.2$\,K.  

\subsection{Detachment of the FM and AFM components}

In order to check whether the anomalous maximum at $T_{ord}$, see Fig.~\ref{F1}a and Fig.~\ref{F2}a, is related to the competition between two different 
magnetic phases  \cite{Martinelli}, one can explore the possibility of subtracting one of these contributions from the total measured signal $\chi_{meas}$ to undress the other. Labeling both 
components as FM: 
$\chi_{FM}(T)$, and AFM: $\chi_{AFM}$, such subtraction can be expressed as $\chi_{AFM}  = \chi_{meas } –  \chi_{FM}$, as schematized in Fig.~\ref{F2}a.

The key information is provided by neutron diffraction measurements \cite{Martinelli} which shows a phase separation between FM ($\approx 33\%$) and AFM ($\approx 66\%$) at 
$T\to 0$, with very different temperature dependencies, see  Fig.~\ref{F2}b. 
As shown in the figure, the FM component is nearly temperature independent below 7K, but it  shows a broad shoulder around $T_{mx}\approx 10$K, with the full signal extrapolating to zero at 
$T\approx 13.8$K. This suggests the presence of two ferromagnetic contributions: one from the pure FM($T$) phase and the other as a spurious (unknown) contribution $L_{10K}(T)$ 
\cite{1-1-2}. 

The respective temperature dependencies of 
$\chi_{FM}(T)= FM(T)+ L_{10K}(T)$, can be described by two simple functions:
\begin{equation}
 FM(T) = \alpha  \tanh [(T_N -T)/1.6]                               
\end{equation}
\begin{equation}  
L_{10K} (T) = \beta [(T-T_{mx})^2 + \sigma^2]
\end{equation}
as depicted in the figure.
The $FM(T)$ function can be associated with the ferromagnetic order parameter $\Psi_{FM}(T)$ 
which is zero for $T>T_N$, while $L_{10K}(T)$ is a Lorentzian function accounting for the broad 
anomaly centered at $T_{mx} \approx 10$K, with a width $\sigma = 1.7$K, being $\beta$ a scaling factor related to the intensity of that magnetic signal.

Since the resulting $\chi_{AFM}(T)$ curve should be related to the AFM order parameter: $\Psi_{AFM} (T)$,  in Fig.~\ref{F2}b we compare (brown curve) this result with the neutron 
scattering counts using an: $a (b - \chi_{AFM})$ formula, with the scale factors $a= 500$\,counts/emu/molOe and $b=5600$ and taking into 
account the fact that the AFM signal tends to zero when $T\to 0$ and increases monotonically with $T$.

One can see in Fig.~\ref{F2}a the result of this subtraction (i.e. $\chi_{AFM}$) which was performed for  the lowest measured field $B_{eff} = 21$\,Oe. This curve can be described by the 
function: $\chi_{AFM}(T,21Oe) = 0.1\, T+0.015\,   T^{2.5}\, e^{-10/ T}$emu/mol\,Oe, which is consistent with the expectations for an anisotropic AFM system \cite{SpinW}. To improve the fit, a 
linear T contribution was added indicating the presence of a continuous distribution of magnetic excitations below 6K and a gap in the magnon spectrum $\Delta (21\,Oe) = 10$\,K.

\begin{figure}
\begin{center}
\includegraphics[width=20pc]{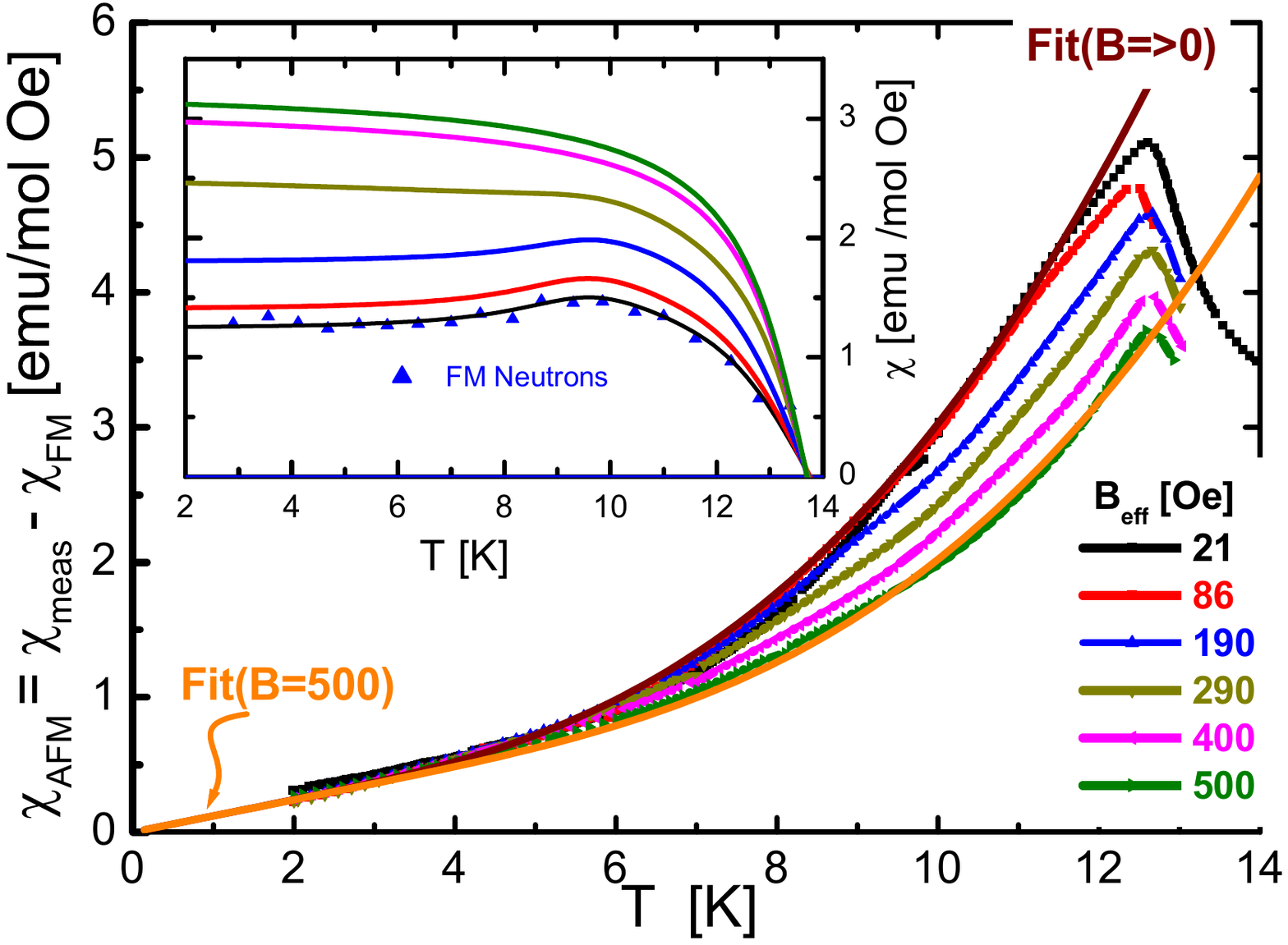}
\caption{(Color online) Temperature dependence of the $\chi_{AFM} (T,B)$ contributions for $0 < B < 500$\,Oe after subtractions of respective $\chi_{FM}(T,B)$ from $\chi_{meas}(T,B)$, see *
text. Inset: Fit of the field dependence $M vs T$ of the FM component in the ordered phase used for the subtraction. Blue triangles and  $\chi_{FM}(T,B=0)$ are from Fig.~\ref{F2}b.
\label{F3}}
\end{center}
\end{figure}

\subsubsection{Field dependence of the FM contribution}

Once the values of parameters $\alpha$ and $\beta$ have been determined in eqs.(1) and (2) for $\chi_{FM}(T,B_{eff} = 0)$, the subtraction procedure can be applied for higher $B_{eff}$ values 
with the same $\chi_{FM}(T)= FM(T)+ L_{10K}(T)$ functions and the corresponding  $\alpha(B)$ and  $\beta(B)$ parameters.  The checking constraint is that the obtained $\chi_{AFM}(T,B)$ data 
have must have the same temperature dependence as that for $\Psi_{AFM} (T)$, represented in Fig.~\ref{F2}b by the AFM counts of neutron diffraction. Note that in the figure the calculated $
\chi_{AFM}(T)$ refers to the right axis with inverted growth since $\chi_{AFM}\to 0$ for $T\to 0$.

\begin{figure}
\begin{center}
\includegraphics[width=20pc]{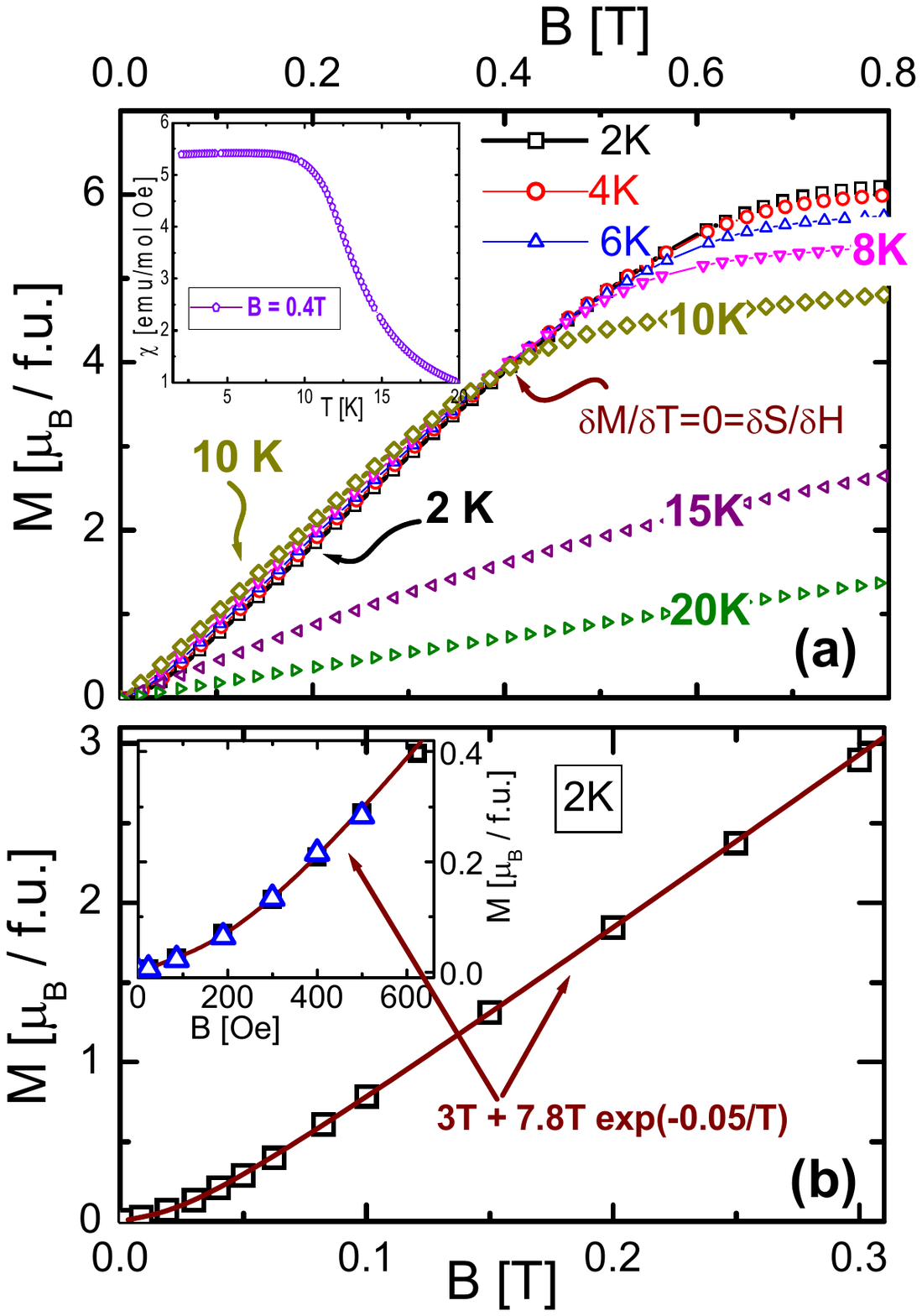}
\caption{(Color online) a) Low field magnetization isotherms in the range $2\leq T\leq 20$\,K, showing the crossing point ($\partial M/\partial T =0$) at $B \approx 0.4$\,T and $T<10$\,K. The 
inset contains the corresponding isopedia at $B=0.4$\,T with a $\chi_{meas} \neq f(T)$ plateau for $T<10$\,K. b) Fit of $M(2K,B)$ up to $B = 0.3$\,T (brown-curve) indicating that, after a weak 
increase, $M(B)$ tends to grow linearly. 
Inset: comparison between the values of $M(B)$ (black squares) and the $\alpha(B)$ coefficient (blue open triangles) after scaling due to different units. The solid brown-curve indicates that the 
change of regime is characterized by a gap  $\Delta = 0.05$\,T.\label{F4}}
\end{center}
\end{figure}

In Fig.~\ref{F3} we show these results for $0<B_{eff}<500$
\,Oe, including the fits for $B_{eff}\to 0$ (brown curve) and 500\,Oe (orange curve). The only change in the $\chi_{AFM}(T,B)$ fit is the enhanced gap that reaches $\Delta(500\,Oe)= 18$\,K. 
While $\alpha(B)$ increases moderately, $\beta(B)$ decreases and tends to zero around $B_{eff} 
\approx 400$\,Oe. The inset of Fig.~\ref{F3} collects the $\Psi_{FM} (T,B)$ curves calculated using the $\alpha(B)$ and $\beta(B)$ coefficients up to 500\,Oe, showing that the $L_{10K}(T)$ contribution disappears around this field.

\subsection{Magnetization} 

Together with the peculiar maximum at $\chi(T_{ord})$, another unusual behavior is observed in this compound. At $T= 2$\,K, the magnetization $M(B,T)$ increases quite linearly up to $B= 
0.4$\,T, see Fig.~\ref{F4}a. At higher fields, $M(B)$ tends to saturate to $M_{sat} = 6.8\,\mu_B$/f.u. after undergoing a pronounced curvature above $B = 0.6$\,T. 

Notably the isotherm $M(B,2K)$ shows slightly lower values than the $M(B, 10K)$ isotherm. This peculiar feature is related to a crossing of these and intermediate isotherms ($T=4$ and 8\,K) at 
a point where $\partial M/\partial T=0$. Two properties can be inferred from this zero value derivative: i) at constant field, it can be written as $\partial \chi /\partial T = 0$ that means a 
constant $\chi(T)$ value at that field, as shown in the inset of Fig.~\ref{F4}a for $B=0.4$\,T. ii) applying the thermodynamic relationship: $\partial M/\partial T = \partial S/\partial H$, this  
means that there is a field independent entropy collected around 10\,K. This peculiarity will be discussed in the section devoted to thermal properties.

A more detailed analysis of $M(2K,B)$ at low field (see Fig.~\ref{F4}b) indicates that, after a weak (AFM-like) increase, $M(B)$ tends to grow linearly according to a heuristic function: 
$3\times T+7.8 \times T\times e^{-0.05/T}$. 
A further check of the validity of the subtraction procedure used in Section B can be done by comparing the variation of the FM-$\alpha(B)$ coefficient (blue open triangles) obtained from the 
$\chi_{FM} (T,B)$ results with the direct measurement of $M(B)$ at 2K (black squares) because $\alpha(B)$ is proportional to the intensity of the FM signal. 

\section{Thermal properties}

\subsection{Specific Heat}

\begin{figure}
\begin{center}
\includegraphics[width=20pc]{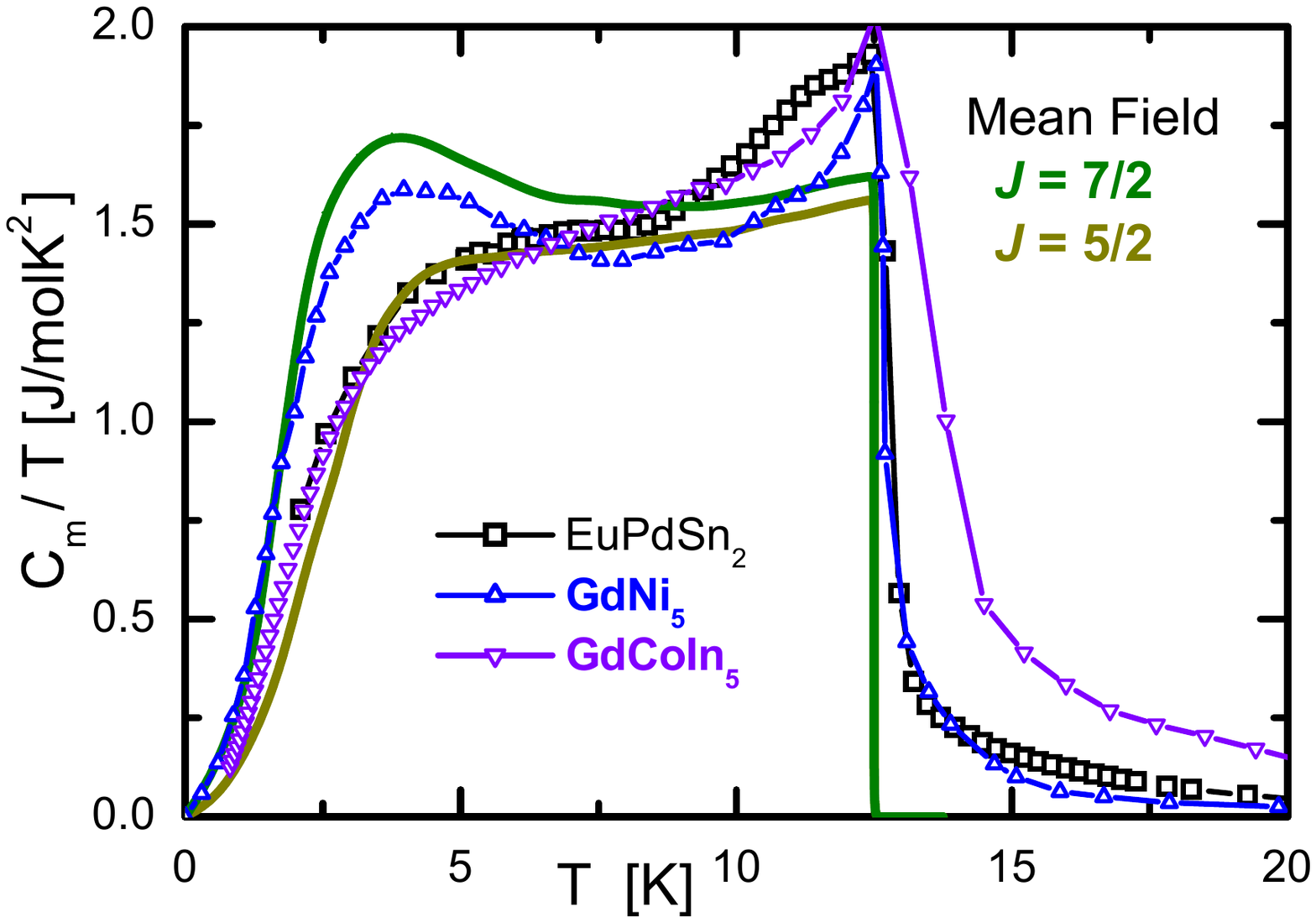}
\caption{(Color online) Comparison of the specific heat behavior of EuPdSn$_2$ with two reference Gd compounds with $J =7/2$: GdNi$_5$ \cite{GdNi5} and GdCoIn$_5$\cite{GdCoIn5}, 
after renormalization of the respective $T_{ord}$. Two (continuous) curves showing the expected behavior for $J=$7/2 and 5/2 specific heat in MFA are also included for comparison.
 \label{F5}}
\end{center}
\end{figure}

Fig.~\ref{F5} shows the temperature dependence of the magnetic contribution to the specific heat $C_m/T(T)$ of EuPdSn$_2$ at zero magnetic field, which looks quite different from the expected 
for a system with $J=7/2$ 
in mean field approximation MFA (green curve) \cite{AmJPhys}. In order to show that  such a  deviation from the predicted behavior is not an isolated case, we included two Gd-based  
compounds that can be taken as a reference for $J =7/2$ atoms. In both cases their respective $T_{ord}$ were renormalized to the $T_{ord}$ of EuPdSn$_2$.
While the $C_m/T(T)$ of GdNi$_5$ \cite{GdNi5} approaches the theoretical prediction, GdCoIn$_5$  \cite{GdCoIn5} behaves very similarly to EuPdSn$_2$. 
Notably, the observed $C_m/T(T)$ behavior of these two compounds is better described by the MFA for $J=5/2$ (dark yellow curve). This means that 
the eight-fold ground state does not split homogeneously (i.e. Zeeman like) within the ordered phase when the internal (or molecular) field increases. In all measurements, $C_m/T(T\to 
T_{ord})$ increases significantly, so that the jump in the specific heat $\Delta C_m$ at $T=T_{ord}$ exceeds the prediction for a $J=7/2$ system in MFA \cite{AmJPhys}. Furthermore, 
in GdCoIn$_5$ \cite{GdCoIn5} the  $C_m/T$ cusp at $T_{ord}$ was associated with a similar cusp in thermal expansion \cite{GdCoIn5}, implying an unusual accumulation of entropy around 
this temperature. 

\begin{figure}
\begin{center}
\includegraphics[width=20pc]{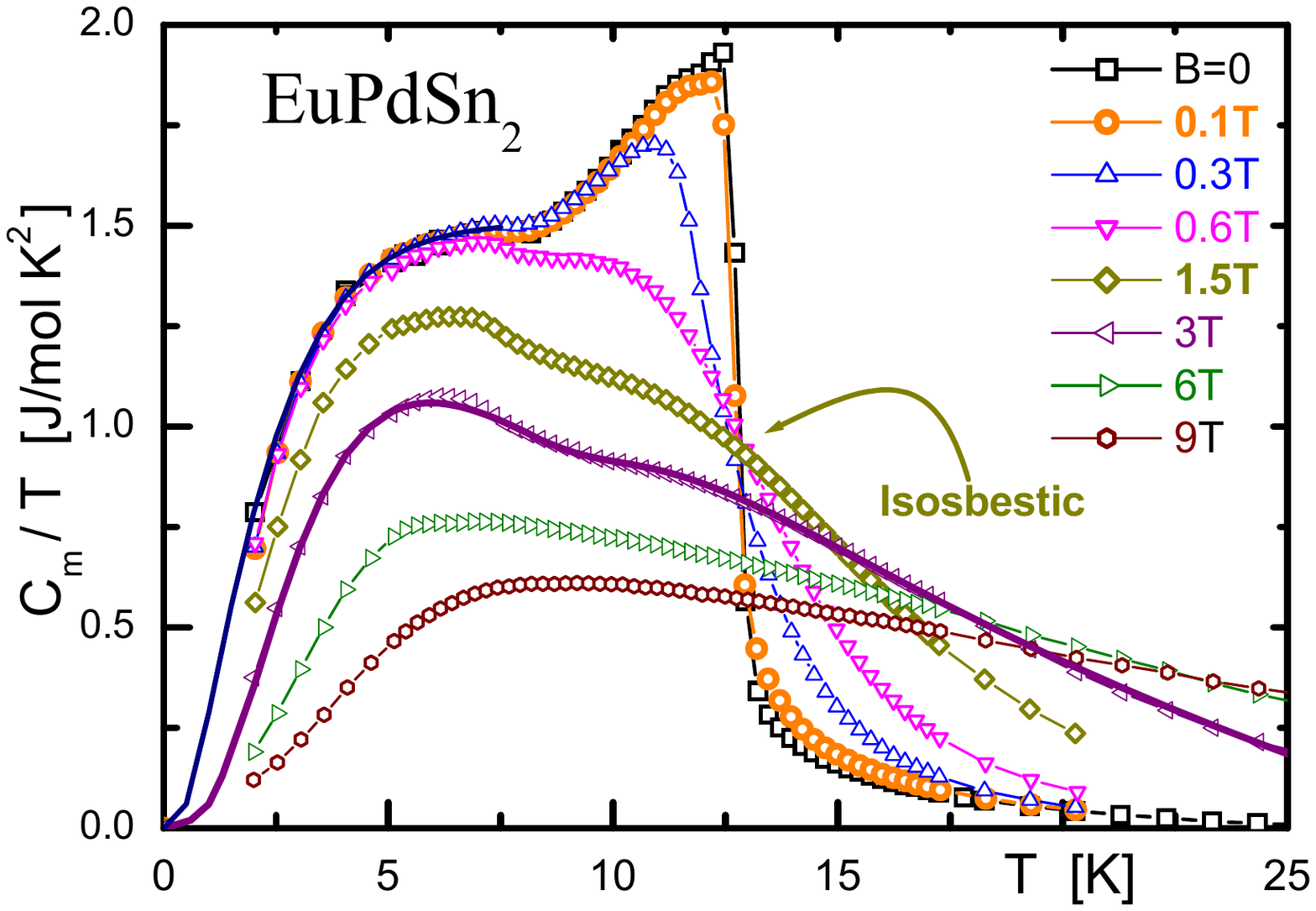}
\caption{(Color online) Specific heat divided temperature, $C_m/T$, for different applied fields. The yellow arrow indicates the isosbestic point observed up to  $B= 1.5$\,T. The navy curve for $B=0$ 
and $T\leq 7$\,K corresponds to a $C_m(T\to 0)$ extrapolation later used for entropy evaluacion, while the violet one represents a fit to the $C_m/T$ results for $B=3$\,T, see text. 
 \label{F6}}
\end{center}
\end{figure}

Under applied magnetic field, $C_m/T(T,B)$ shows the typical variation for a FM behavior, see Fig.~\ref{F6}, with the exception of two nearly independent humps. While one remains practically 
unchanged around 6K up to $B\approx 3$\,T the intensity of other, which has its origin in the $T_{ord}$ cusp, decreases significantly its intensity with field. One can recognize the presence of 
an isosbestic point where $C_m/T \neq f(B)$ \cite{isosb}, that holds at least 
up to 1.5\,T. This means that above this field, $B$ overcomes the internal field and increasingly  dominates the scenario.  

Fig.~\ref{F6} also shows a fit of the $C_m/T(T)$ data for $B=3$\,T. This fit, which includes the two observed humps, is performed using two contributions based on the relation 
$C_J(x)/R= x^2 \partial B_J/\partial x$ \cite{AmJPhys}, where $B_J(x)$ is a Brillouin function with $x=g_J \mu_B J B/k_B T$, and $J$ the main quantum number used for the fit. 
\begin{equation}
C_J(x)/R=[A \times csch(A)]^2 - [B \times csch(B)]^2
\end{equation}
where $A=x/2J$ and $B= (2J+1)x/2J$.
In this case, the contribution with the maximum at $T\approx 6$\,K is represented by $C_{3/2}$ and involves $N=2J+1=4$ levels, while the other: $C_{1/2}$, describes the access to an 
equivalent set of higher excited 
states with a maximum around $T \approx 12$\,K. This $B=3$\,T isopedia was chosen because, as already mentioned, once the isosbestic point is overcome by the external field $B$, 
the energy spectrum of the magnetic levels is increasingly less affected by the anisotropic internal field.

The use of Brillouin functions for this description is a rather simplified approximation. Nevertheless, this choice is supported by the good fit obtained, see  Fig.~\ref{F6}, and confirmed 
by the entropy involved: $\Delta S_m = Rln(8)$. Remind that the 
total entropy for $J=7/2$ is $\Delta S_m = Rln(4) + Rln(2) =Rln(8)$, where the first term correspondsto $C_{3/2}$ and the second to $C_{1/2}$. This analysis confirms that the eight 
fold ground state of $Eu^{2+}$ is split into at least two sets of four levels. 

\subsection{Entropy}

\begin{figure}
\begin{center}
\includegraphics[width=20pc]{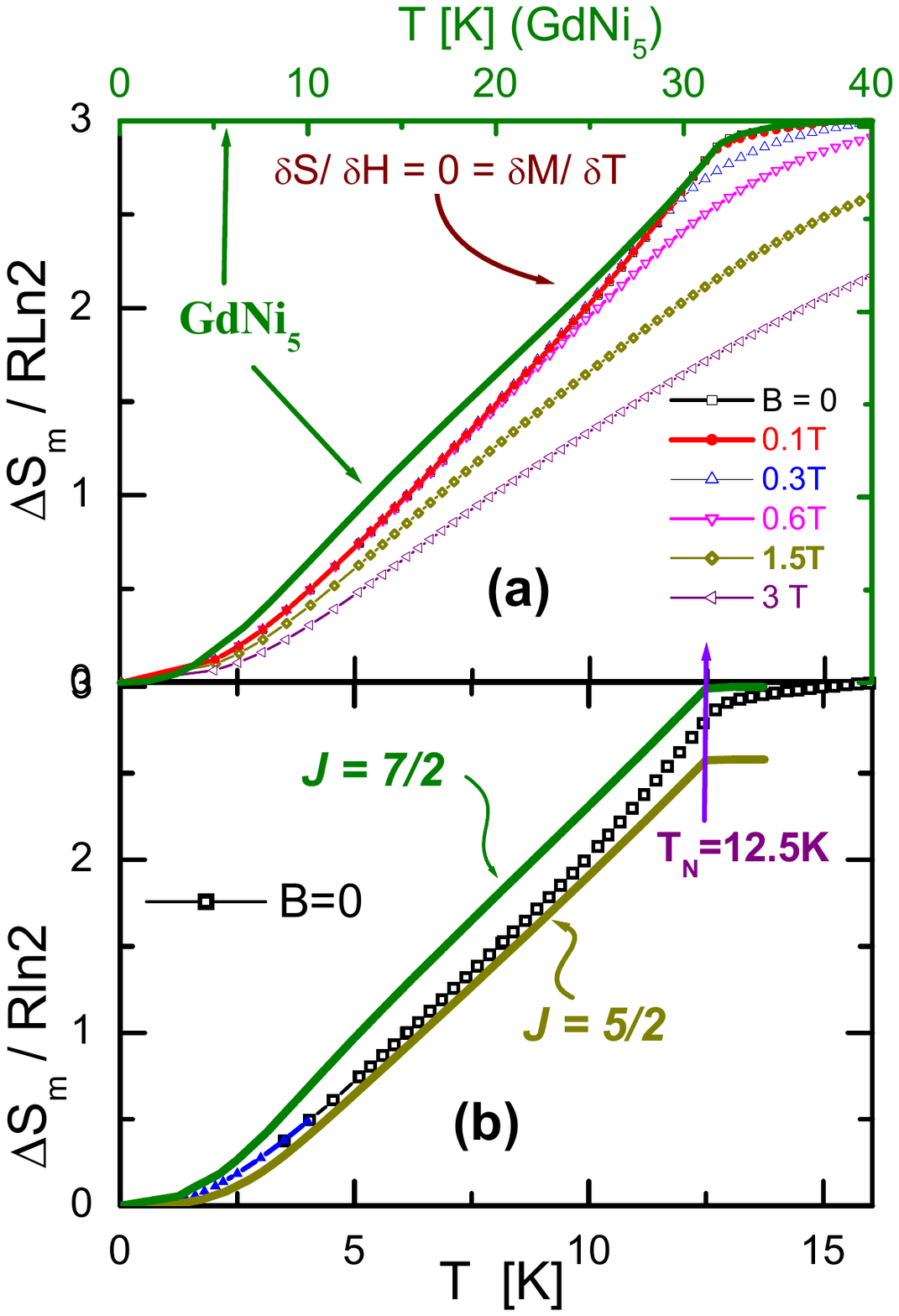}
\caption{(Color online) a) Magnetic entropy $\Delta S_m(T)$ gain for different fields normalized to R$\ln(2)$, and comparison with the reference compound GdNi$_5$ for $J=7/2$ (continuous 
green curve using the upper temperature axis). The crossing point at $T=10$\,K and $B\leq 0.4$\,T is indicated as $\partial S/ \partial H =0$. b) Comparison of  $\Delta S_m(T)$ for EuPdSn$_2$ 
with values for $J=7/2$ and 5/2 computed from MFA theory. The blue curve for $T\leq 4$\,K 
corresponds to the $\Delta S_m(T\to 0)$ extrapolation. \label{F7}}
\end{center}
\end{figure}

Although the total entropy collected at $T=T_{ord}$ corresponds to the expected $\Delta S_m = R\ln8$, between 0 and ~10K (see Fig.~\ref{F7}a) the $\Delta S_m(T)$  trajectory is well below 
that for a $J = 7/2$ system like GdNi$_5$, but closer to that calculated for $J=5/2$ (see Fig.~\ref{F7}b) untill it increases significantly at $T\to T_{ord}$. This confirms that the energies 
of the $N = 2J+1 = 8$ levels are not monotonically distributed as expected for a standard Zeeman type splitting as previously quoted. 

According to the $\partial M/\partial T = 0 =\partial S/\partial H$ relation, $\Delta S_m(T)$ curves are superimposed up to 10\,K and fields up to 0.3\,T as shown in Fig.~\ref{F7}a, in 
agreement with the constant entropy character of this $T$ and $B$ region. As expected, the increase of entropy observed above 10\,K, see  Fig.~\ref{F7}b, includes the difference between 
$J=5/2$ and $7/2$ predictions. This kind of recovery of entropy to 
reach the $Rln(8)$ value also indicates a relative accumulation of magnetic excitations close to $T_{ord}$. In fact, this entropy difference $\Delta S_{T_N} = 1.5$\,J/molK, see  Fig.~\ref{F7}b, 
is close to the 
difference between $N=8$ degeneracy of $J=7/2$ and $N=6$ of $J=5/2$: $\Delta S_{T_N} =  
2/3 R\ln(8/6)= 1.58$ J/molK. 

It should be noted that this distribution of energy levels at $B=0$ is different from that obtained  from the fit at $B=3$\,T. This difference can be attributed to the fact that, at $B=0$ the effect of  
the internal field anisotropy dominates, whereas at $B=3$\,T (above the 
isosbestic point) the dominant magnetic symmetry corresponds to that of the applied field acting  randomly on the grains of a polycrystalline sample directions. 

\section{Magnetic Phase Diagram}

\begin{figure}
\begin{center}
\includegraphics[width=20pc]{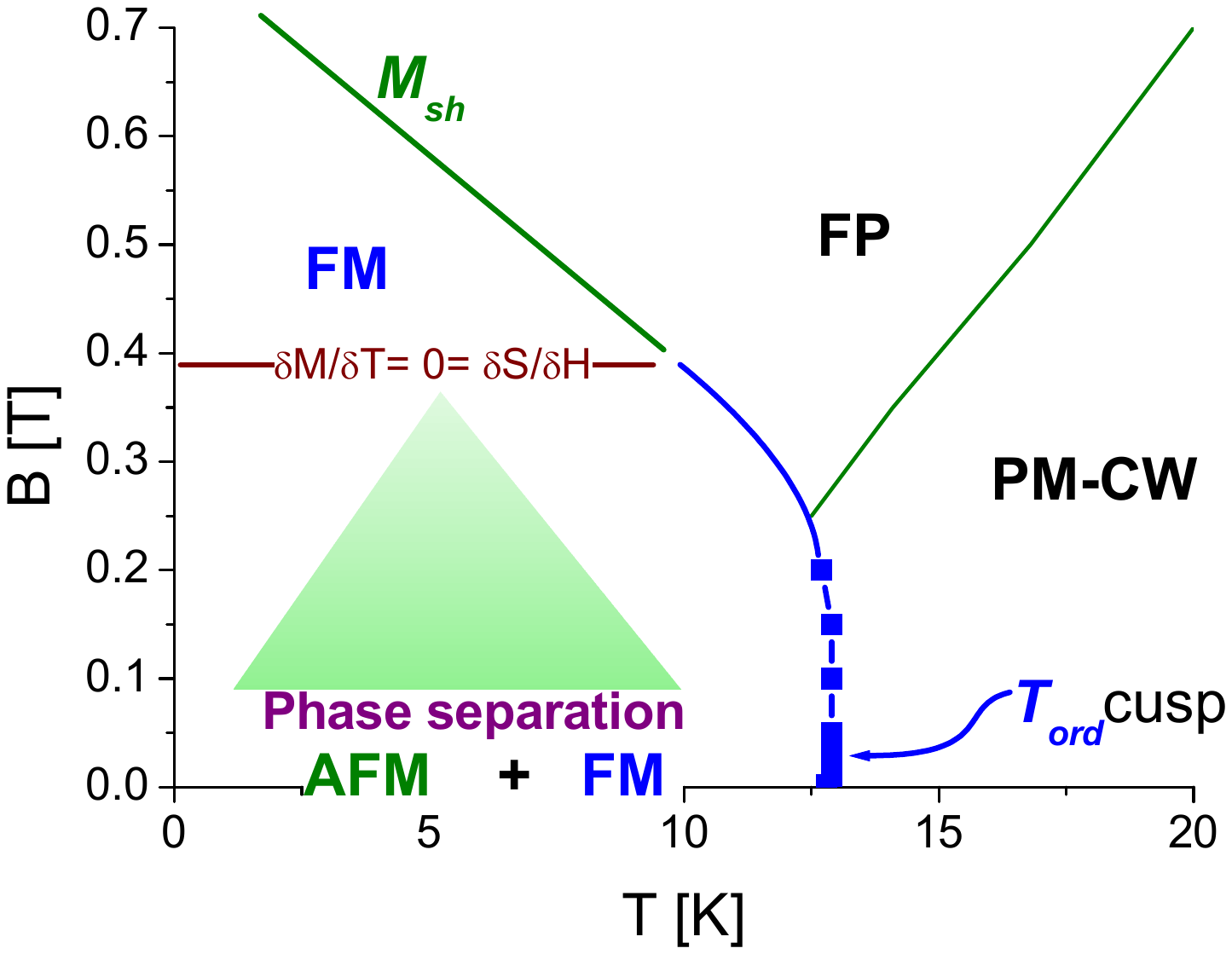}
\caption{(Color online) Tentative magnetic phase diagram for EuPdSn$_2$. $M_{sh}$ indicate the evolution of a shoulder in $M(B)$ where the linear increase turns towards saturation. The green 
triangle indicates how the intensity of the AFM component decreases by increasing external field. PM-CW indicates the region where the Curie-Wiess law is observed, and FP where deviation  from 
that behavior occurs due to field polarization of the magnetic moments.
\label{F8}}
\end{center}
\end{figure}

In this compound, the only phase transition observed is that between the paramagnetic PM   
and the ordered phase at $T=T_{ord}$, which vanishes around $B \approx 0.2$\,T, see Fig.~\ref{F8}. At higher fields it can be traced as a shoulder in $M(B)$ (see Fig.~\ref{F4}a)   
labeled $M_{sh}$, where the linear increase turns towards saturation. 
The green triangle in Fig.~\ref{F8} indicates how the intensity of the AFM component decreases with increasing field, from about 60\% (at $B = 0$) of the ordered phase to zero at $B 
\approx 0.4$. At this field, the 
thermodynamic relation: $\partial M/ \partial T = 0 =  \partial S/ \partial H$ is confirmed  between $0 < T < 10$\,K in the magnetic and thermal properties. Above this field, the ground 
state mostly behaves as FM, and above $T_{ord}$ one sees the crossover from the Curie-Weiss law (PM-CW) to a field polarized (FP) region.

\section{Conclusions}

We have seen that the FM and AFM contributions can be quantitatively separated from the magnetic susceptibility signal by taking the temperature dependence of the neutron counts 
collected for the FM component as a reference. Though this result does not provide information on  the spatial distribution of these phases, the simple additive character of both signals suggests 
that it is not an 'in situ' coexistence, rather an independent distribution in the volume of two different order parameters. As it can be deduced from the magnetic susceptibility and the  
magnetization behavior 
under magnetic field, the FM phase rapidly overcomes the AFM phase 

The anomalous evolution of the magnetization with temperature between 2 and 10\,K shown in  Fig.~\ref{F4}a, together with that of the entropy: $\Delta S_m(T)$ for $0 \leq B \leq 0.3$\,T, see 
Fig.~\ref{F6}a, indicate the increasing presence of the AFM phase at low temperatures and low fields. This peculiar situation provides the conditions for an experimental verification of 
the thermodynamic ralation $\partial M/ \partial T = 0 =  \partial S/ \partial H$, which is  confirmed by  the temperature independent susceptibility at $B \approx 0.4$\,T included in the 
inset of Fig.~\ref{F4}a. 

Another unexpected feature of this compound is the non-mean field behavior of $C_m/T(T)$, Fig.~\ref{F5}. This dependence does not correspond to a Zeeman energy 
distribution of the eight fold ground state of the Eu$^{2+}$ atoms, which are split by the internal molecular field. Notably, at zero field the data are better described by the 
$J=5/2$ case than by the $J=7/2$. However at $B=3$\,T, which is stronger than the molecular field, the better approach is the one with two sets of quartets.

Strictly speaking, this scenario is not only observed in this Eu$^{2+}$ compound, but also in GdCoIn$_5$, which is included in  Fig.~\ref{F5} for comparison. In all cases, the $L=0$ 
character of the orbital number inhibits pointing to a particular electronic configuration for this  anomalous behavior. However, since magnetic anisotropy characterizes these 
compounds, the non-isotropic magnetic environment may be responsible for the observed distortion.  Thus, the anisotropic distribution of magnetic interactions with neighboring magnetic atoms 
may explain the failure of an isotropic hypothesis such as the MFA theory. Note that the degeneracy of the $N=8$ ground state recovered as soon as the system approaches the intrinsically isotropic 
paramagnetic state at $T_{ord}$, in a temperature range where the anisotropic internal field strongly decreases. A better knowledge of the role of the magnetic interactions in an anisotropic 
environment requires a sophisticated mathematical analysis, which eis beyond the scope of this work.

\end{document}